\documentclass[twocolumn,aps,showpacs,preprintnumbers,superscriptaddress,amsmath,amssymb]{revtex4}
\usepackage{graphicx,epsf,epsfig}
\usepackage{dcolumn}
\usepackage{bm}
\usepackage{color}
\usepackage{ulem}
\usepackage{subfigure}
\begin{document}
\title{Rectification by charging -- the physics of contact-induced current asymmetry
in molecular conductors}
\author{O. D. Miller}
\affiliation{Dept. of Electrical and Computer Engineering, University of Virginia}
\author{B. Muralidharan}
\affiliation{School of Electrical and Computer Engineering, Purdue University}
\author{N. Kapur}
\affiliation{Dept. of Chemical Engineering, University of Virginia\\
}
\author{A. W. Ghosh}
\affiliation{Dept. of Electrical and Computer Engineering, University of Virginia}

\date{\today}
\begin{abstract}
We outline the qualitatively different physics behind
charging-induced current asymmetries in molecular conductors
operating in the weakly interacting self-consistent field (SCF) and
the strongly interacting Coulomb Blockade (CB) regimes. A
conductance asymmetry arises in SCF because of the unequal
mean-field potentials that shift a closed-shell conducting level
differently for positive and negative bias. A very different current
asymmetry arises for CB due to the unequal number of open-shell
excitation channels at opposite bias voltages. The CB regime,
dominated by single charge effects, typically requires a
computationally demanding many-electron or Fock space description.
However, our analysis of molecular Coulomb Blockade measurements
reveals that many novel signatures can be explained using a
{{simpler}} orthodox model that involves an incoherent sum of Fock
space excitations and {\it{hence treats the molecule as a metallic dot or an island}}.
This also reduces the complexity of the Fock space description by just including various charge configurations only,
thus partially underscoring the importance of electronic structure,
while retaining the essence of the single charge nature of the transport process.
We finally point out, however, that the inclusion of electronic structure and hence well-resolved Fock space excitations is
crucial in some notable examples.
\end{abstract}
\maketitle
\section{Introduction}

\begin{figure}[ht]
\centerline{\epsfig{figure=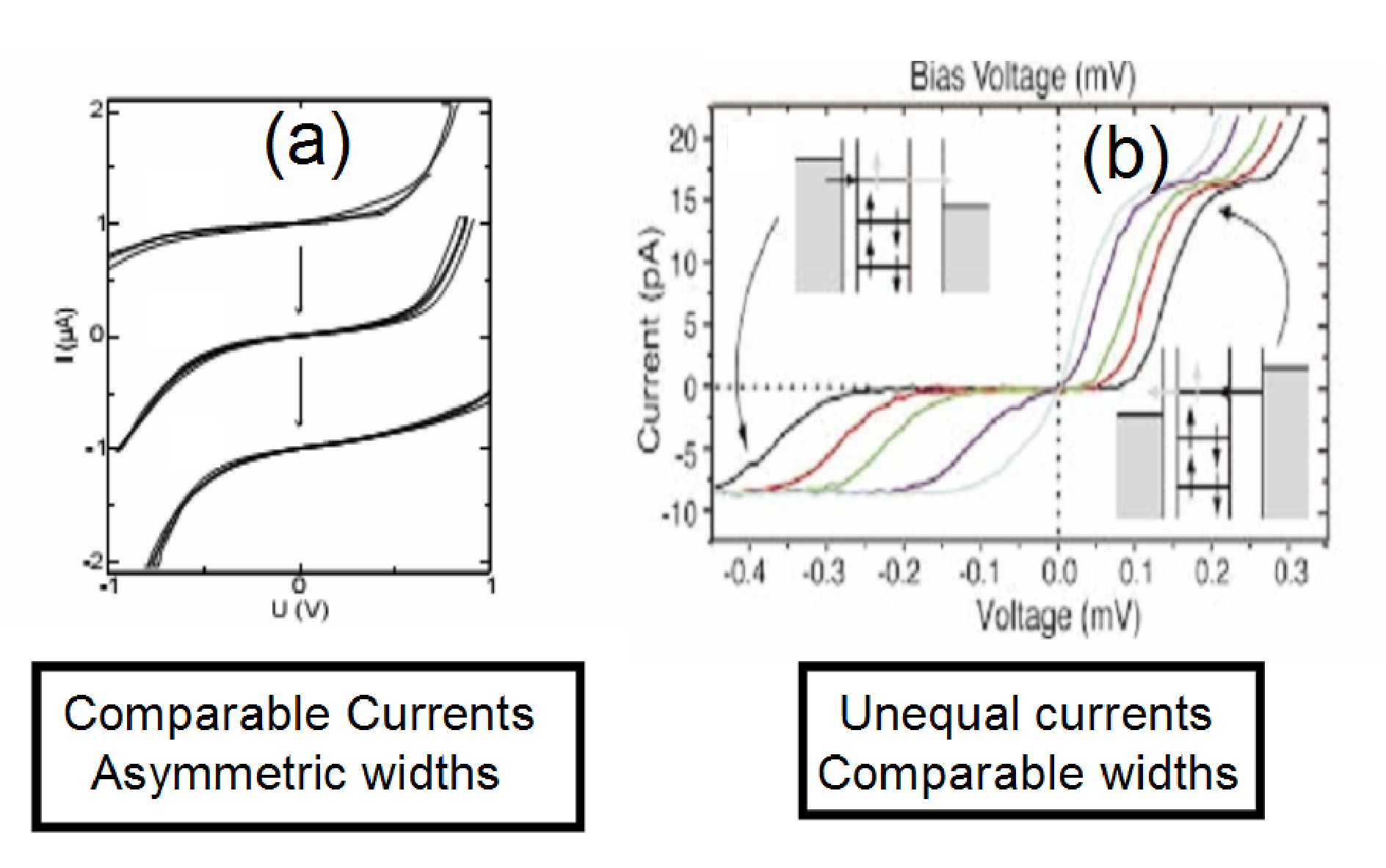,width=3.0in}}
\caption{(Color online) Experiments showing (a) comparable currents reached over unequal
voltage widths \cite{rei1} in the SCF limit; (b) unequal currents with comparable
widths in the CB limit \cite{rralph2}.} \label{f0}
\end{figure}

Ever since its inception \cite{avi}, molecular rectification continues
to be of great practical interest. While rectification could arise
from asymmetries in the intrinsic molecular structure or vacuum 
barriers at the ends, there are multiple experiments \cite{rei1,jpark,rscott}
that exhibit pronounced asymmetries in current voltage (I-V) or
conductance voltage (G-V) characteristics due to unequal coupling
with contacts. Fig.~\ref{f0} shows that the nature of contact-induced
asymmetry is qualitatively different depending on the nature of the
molecule-contact bonding. For molecules strongly coupled with the
contacts with adiabatic charge addition, equal current plateaus are
reached over unequal SCF voltage widths (Fig.~\ref{f0}a), leading
to prominent conductance asymmetries \cite{rei1}. The origin of this
asymmetry is the different average charging energies that generate {\it{
unequal
mean-field potentials}} for opposite bias voltages \cite{rasymm}.
Reducing the contact-molecular coupling drives the system into CB, where the
intermediate open-shell
current values are also asymmetric \cite{rralph2} (Fig.~\ref{f0}b). This asymmetry
has a different physical origin rooted in its many-body excitations, driven
by the {\it{unequal number of discrete charge
addition and removal channels}} at opposite bias. It is thus clear that
the physics of rectification can differ widely depending on
the strength of the electron-electron interaction.

The non-equilibrium Green's function (NEGF) formalism is widely
established for treating quantum transport in the SCF regime, for a
whole variety of materials from nanoscale silicon transistors to
molecules, nanowires, nanotubes and spintronic elements. The ability
to incorporate sophisticated quantum chemical models
\cite{max_dft,damle2} through averaged potentials makes the NEGF-SCF
scheme particularly attractive in the community. What is not widely
appreciated is that this approach does not readily translate to the
CB regime, even qualitatively \cite{rbhasko,rbhaskob,rbhaskoc}. The
CB regime, observed in molecules with weak contact coupling
\cite{jpark,rscott}, manifests clear signatures of single-electron
charging such as suppressed zero-bias conductances and abrupt jumps
in current. Although several approximate treatments
\cite{max_dft_2,rpal,ssan,pals} have been suggested to handle CB
within effective one-electron potentials, the inherent difficulty
arises from the fact that the open-shell current levels depend on
full exclusion statistics in its many-body Fock space. Even for a
minimal single-orbital model, it is easy to establish that while the
open-shell current plateau widths depend on the correlation
strengths, their heights are independent of correlation and, in that
sense, universal \cite{rbhasko}. The transport problem in CB maps
onto a rather difficult combinatorial problem in many-body space
that is hard to capture {\it{a-priori}} through a one-particle SCF
potential, or improve upon phenomenologically.

It seems likely that in the limit of weak coupling to contacts a
proper treatment of the above excitations will require solving a set
of master equations directly in the Fock space of the molecular
many-body Hamiltonian \cite{rbhasko, rhettler}. A significant
penalty is the increased computational cost that requires
sacrificing the quantum chemical sophistication of ab-initio models
in lieu of an exact treatment of the Coulomb interaction in simpler,
phenomenological models. Within such an exactly diagonalizable
model, one can capture transport features quite novel and unique to
the CB regime, such as inelastic cotunneling, gate-modulated current
rectification and Pauli spin blockade
\cite{rbhasko2,rbhasko,rsiddiqui}. The presence of contact asymmetry
makes these features even more intriguing, while somewhat
simplifying the analysis by effectively driving the system into
equilibrium with the stronger contact.

In this paper, we first identify the origin of current asymmetry
with a minimal system of a single spin degenerate energy doublet,
employing NEGF in the SCF limit and master equations for sequential
tunneling \cite{rralph} in the CB limit (we refer to this as a
``Fock-space" CB model). We then extend this ``Fock-space model" to
a general molecular Hamiltonian to explain how multiple orbitals in
CB allow simultaneous sequential tunneling into excited states,
making the conductance peak heights vary with gate voltage. We also
explain the origin of exchange in conductance peak asymmetry by the
neutral and singly charged molecule. While this Fock space model
involves a computationally intensive, exactly diagonalized many-body
Hamiltonian about the charge degeneracy point arising from the
different excitation spectra accessed, in many cases a simpler
approximation works. Our analysis of the experimental trends in the
CB regime reveals that many of the novel signatures such as gate
dependence of conductance peaks and asymmetry flipping can be
explained using a {\it{simpler}} ``orthodox model" that involves an
incoherent sum of Fock space excitations. We conclude by pointing
out some limitations of such a simplified approach, as well as
possible applications that may need careful attention to the
detailed excitation spectrum.

\section{Origin of Current Asymmetries -- The essential physics}
As mentioned earlier, there are two distinct physical limits of transport. In the
SCF limit, contact broadenings $\Gamma$ are greater than or comparable with the
single electron charging $U$. In the opposite CB limit $U\gg\Gamma$ and single-electron
charging dominates. Conductance asymmetries in both regimes
of transport have been experimentally observed in molecular conduction. While there are
ways to handle each regime separately, treatments are inherently perturbative, with an
approximate treatment of correlation (in terms of $U/\Gamma$) for the SCF regime, and
an approximate treatment of broadening (in terms of $\Gamma/U$) for the CB regime. The
lack of a small parameter in the intermediate coupling regime ($U \sim \Gamma$) makes
the exact treatment of transport, even for a simple model system, potentially
intractable \cite{rbhaskob}.

\begin{figure}
\centerline{\epsfig{figure=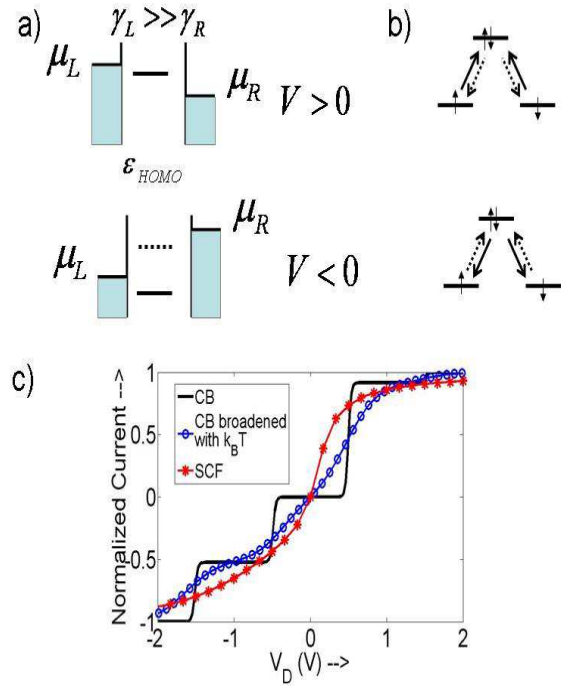,width=3.8in,height=3.8in}}
\caption{(Color Online) Origin of Current Asymmetries: a) In the SCF
regime (stars), when $\gamma_L \gg\gamma_R$ for $V>0$, the left
contact keeps the HOMO $\epsilon_{HOMO}$ level filled to its neutral
charge making conduction possible, while for $V<0$, the left contact
empties the electron leading to the level floating out of the bias
window, creating {\it{unequal plateau widths}} for opposite bias. b)
In the CB regime (bold solid line) however, for $V<0$, charge
removal process by the right contact is rate determining and occurs
in two different ways, while for $V>0$, charge addition process is
rate determining and occurs in only one way. This leads to
{\it{asymmetry in plateau heights}} along the two bias directions.
c) While the transition between the two limits is hard to accomplish
correctly, a phenomenological broadening through an artificial
enhanced temperature (circles) illustrates how the open-shell
intermediate CB plateau morphs into a higher effective broadening,
restoring the SCF result in the limit of $\Gamma \gg U$} \label{f1}
\end{figure}

The origin of asymmetric I-Vs can be easily elucidated with a
minimal model for current conduction through a spin degenerate,
filled (closed-shell) molecular level doublet. We assume equal
{\it{capacitive}} couplings but unequal {\it{resistive}} couplings
to the contacts, so that the molecular level shifts by half the
applied Laplace potential and the current onsets arise symmetrically
around zero bias. In the SCF limit, contact asymmetry results in
equal currents adiabatically smeared out over a larger voltage width
along one bias direction than the other. This charging based
asymmetry has been experimentally seen \cite{rei1}, and can be
intuitively rationalized as follows. Consider a spin degenerate
energy level, a highest occupied molecular orbital (HOMO) for
example, that is fully occupied at equilibrium. For asymmetric
contact couplings $\gamma_L \gg \gamma_R$, charge addition dominates
for positive bias on the right contact and removal for negative
bias, as shown in Fig.~\ref{f1}(a). For positive bias the energy
level is maintained at neutrality by the dominant left contact and
the current flow through the level is determined by the removal
rate. Along reverse bias, in contrast, charge removal by the left
contact drives the system away from neutrality towards a net
positive charge, whose Coulomb cost floats the level out of the bias
window. This means that a larger bias is needed to fully conduct
through the level, dragging out the I-V in that direction. The
direction of the asymmetry flips if conduction is through the lowest
unoccupied molecular orbital (LUMO) instead. Notably the peak
currents and initial plateau onsets remain the same along both
directions, but their complete saturations are delayed by different
amounts.

The origin and manifestation of current asymmetry is qualitatively different
in the CB limit, where charge addition or removal is abrupt and in
integer amounts. Given asymmetric contact couplings ($\gamma_L\gg\gamma_R$),
the left contact adds (removes) an electron as soon as the right
contact removes (adds) it, so that the rate determining step becomes
the dynamics of the weaker right contact. For positive bias, charge
removal can happen in two ways, from $\uparrow\downarrow$ to
$\uparrow$ and $\downarrow$, while for opposite bias the right
contact can add a spin in only one way, either $\uparrow$ or
$\downarrow$ to $\uparrow \downarrow$. This scheme of charge
transfer (Fig.~\ref{f1}b) leads to twice the current step for
positive bias than negative \cite{rralph,rralph2}.

An important question is whether one can smoothly transition from
the CB to the SCF asymmetry by progressively increasing the
broadening. While this is hard to do exactly owing to the inherent difficulty involved in
the broadening many-particle states \cite{rWacker}, for the purpose of
illustration one can add various degrees of approximate broadening
\cite{rappenzeller}. We choose to do this by increasing the
temperature and incorporating this through Boltzmann factors in the
many-body occupancies \cite{rgerhard}. As seen in Fig.~\ref{f1}(c),
this approximate treatment morphs the CB asymmetry into the very
different version seen for the SCF limit. {For negative bias on the
weaker contact, `shell-filling' \cite{rzung} of the HOMO level with a net
positive charge creates a CB plateau that is missing in its positive
bias `shell-tunneling' counterpart. This extra CB plateau gets
encased by the broadened manifold, leading to the postponed
conduction seen in the SCF limit. It is worth mentioning though that
the correspondence is only qualitative and is observed to worsen for
higher onset voltages, underscoring the inadequacy of thermal
effects and possibly other phenomenological ways to incorporate
broadening in correlated systems, particularly in the
near equal coupling, non-equilibrium limit which combines shell 
tunneling with shell filling.

\section{Coulomb Blockade Formalism: Fock-space vs Orthodox}
In this paper, we focus mainly on the CB regime. The object of this
section is to present the formalism in detail, beginning with the
one that employs the Fock space. Here, one needs to keep track not
only of various ground state charge configurations, but also various
excitations within each charge state. Following that we derive the
orthodox model that integrates out the effect of excitations in an
incoherent way. A discussion of merits and de-merits of such a
simplification comprises the subsequent sections.
\subsection{Fock-Space Master Equation}
The starting point is a molecular many-body Hamiltonian
\begin{eqnarray}
\hat{H} &=& \sum_{m} \epsilon_{m} n_{m}
+ \sum_{m \neq p} t_{m p} c_{m}^{\dagger} {c_p} \nonumber\\
&+& \sum_{m} U_{m m} n_{m\uparrow} n_{m\downarrow}
+ \frac{1}{2} \sum_{m \neq p} U_{m p} n_{m} n_{p},
\label{eq:mbh}
\end{eqnarray}
where $m, p$ denote the spin-charge basis functions within
a tight binding formulation, with $\epsilon, t$ and $U$ denoting
on-site, hopping and charging terms, respectively. Exactly
diagonalizing this Hamiltonian yields a large spectrum of closely
spaced excitations in every charged molecular configuration. The
lead molecule exchange processes are accounted for within the
sequential tunneling approximation \cite{timm,rbraig}. This results
in a set of master equation that involve transition rates
$R_{(N,i)\rightarrow(N\pm 1,j)}$ between states differing by a
single electron. If one neglects off-diagonal coherences, the master
equation \cite{rbraig} is cast in terms of the occupation
probabilities $P^N_i$ of each N electron many-body state
$|N,i\rangle$ with total energy $E^N_i$. The end result is a set of
independent equations defined by the size of the Fock space
\cite{rralph}
\begin{equation}
\frac{dP^N_i}{dt} =-\sum_j\left[R_{(N,i)\rightarrow(N\pm 1,j)}P^N_i - R_{(N\pm
1,j)\rightarrow(N,i)}P^{N\pm 1}_j\right]
\label{ebeenakker}
\end{equation}
along with the normalization equation $\sum_{i,N}P^N_i =
1$. For weakly coupled dispersionless contacts, parameterized using
bare-electron tunneling rates $\gamma_{\alpha}$ ($\alpha$: left/right
contact) within a Golden Rule treatment, we define the transition-resolved
rate constants \begin{eqnarray}
\Gamma_{ij\alpha}^{Nr} &=& \gamma_\alpha|\langle
N-1,j|c^{}_\alpha|N,i\rangle|^2\nonumber\\ \Gamma_{ij\alpha}^{Na}
&=& \gamma_\alpha|\langle N+1,j|c^{\dagger}_\alpha|N,i\rangle|^2,
\end{eqnarray} where $c^\dagger_\alpha,c^{}_\alpha$ are the
creation/annihilation operators for an electron on the left or right
molecular end atom coupled with the corresponding electrode. The
transition rates are given by
\begin{eqnarray}
R_{(N,i)\rightarrow(N-1,j)} &=& \sum_{\alpha=L,R}\Gamma_{ij\alpha}^{Nr}\left[1-f(\epsilon^{Nr}_{ij}-\mu_\alpha)\right]
\nonumber\\
R_{(N-1,j)\rightarrow(N,i)} &=& \sum_{\alpha=L,R}\Gamma_{ij\alpha}^{Nr}f(\epsilon^{Nr}_{ij}-\mu_\alpha).
\label{rates}
\end{eqnarray}
for the removal levels $(N,i \rightarrow N-1,j)$, and replacing $(r
\rightarrow a,   f \rightarrow 1-f)$ for the addition levels $(N,i
\rightarrow N+1,j)$. $\mu_\alpha$ are the contact electrochemical
potentials, $f$ is the corresponding Fermi function, with single
particle removal and addition transport channels $\epsilon^{Nr}_{ij}
= E^N_i - E^{N -1}_j$, and $\epsilon^{Na}_{ij} = E^{N+1}_j - E^N_i$.
Finally, the steady-state solution to Eq.~(\ref{ebeenakker}) is used
to get the left terminal current
\begin{equation}
I =\pm\frac{e}{\hbar}\sum_{N,ij}\left[R^L_{(N,i)\rightarrow(N\pm
1,j)} P^N_i - R^L_{(N\pm 1, j)\rightarrow(N,i)}P^{N\pm 1}_j \right]
\label{curr}
\end{equation}
where $R^L$ includes the contributions to $R$ from the left contact
alone. In the equation above, states corresponding to a removal of
electrons by the left electrode involve a negative sign. We usually
calculate current in a
break-junction configuration with equal electrostatic coupling with
the leads, $\mu_{L,R} = E_F \mp eV_d/2$.
\subsection{The Orthodox Model}
In the previous sub-section, the computational complexity of the
master equation defined in Eq.~(\ref{ebeenakker}) arises from the
need to keep track of not only charge $N$, but also all
configurational degrees of freedom $i$. Thus the evaluation of
transition rates defined in Eq.~(\ref{rates}) and eventually current
(Eq.~\ref{curr}) depends on our knowledge of various many-electron
wavefunctions $|N, i \rangle$ and total energies $E^{N}_{i}$. The
``orthodox'' theory of single-electron tunneling, however, does not
distinguish between excitation levels \cite{rsold,rkorot}. Note also
that in the orthodox theory, the junctions are generally denoted by
$(1,2)$ instead of $(L,R)$. Eq.~(\ref{curr}) then becomes
\begin{eqnarray}
I & = &
\pm\frac{e}{\hbar}\sum_{N}\left[R_{N\rightarrow{N\pm1}}^{(1)}-
R_{N\rightarrow{N\mp1}}^{(1)}\right] P^{N} \\
& = & \pm\frac{e}{\hbar}\sum_{N}\left[R_{N\rightarrow{N\mp1}}^{(2)}-
R_{N\rightarrow{N\pm1}}^{(2)}\right] P^{N} \label{eq:curr_master}
\end{eqnarray}
We have simplified the second term in the summation using a simple
change of variables.  Let us assume incoherent processes that
introduce an additional exclusion term in the transition process
(appendix). Then a ``golden-rule'' calculation gives
\begin{eqnarray}
&R_{N\rightarrow{N+1}}^{(2)}&=\int_{-\infty}^{\infty}{\displaystyle{2\pi}}\Gamma_{2}^{Na}
D_{2}(E-\mu_{2})f(E-\mu_{2})\nonumber\\
&\times& D_{m}(E-\mu)\left[1-f(E-\mu)\right]dE
\mbox{, } \label{goldenrule}
\end{eqnarray}
where $D_{2}(E)$ and $D_{m}(E)$ are the densities of states of the
right and middle electrodes, and $\mu$ is the Fermi energy of the
molecular system.  If the level separation between one-electron
levels is very small, like in a metallic quantum dot, the densities
of states are approximately constant for the calculation of the
removal and addition rates. By additionally assuming
$\Gamma_{\alpha}^{Na}$ is energy-independent, Eq.~\eqref{goldenrule}
simplifies:
\begin{equation}
R_{N\rightarrow{N\pm1}}^{(j)} =\frac{\hbar}{R_{j}e^{2}} \left(
\frac{-\Delta
    E_{j}^{\pm}}{1-exp(\Delta E_{j}^{\pm} / kT)} \right) \mbox{, }
\label{gamma}
\end{equation}
where $R_j$ is the junction resistance. $\Delta E_{j}^{\pm}$ is the
transition energy corresponding to adding or removing an electron at
the $j^{th}$ contact; from simple electrostatics,
\begin{eqnarray}
\Delta E_{1}^{\pm} = \Delta U^{\pm} \mp \frac{eC_2}{C_{\Sigma}}
            V_D \mp \frac{eC_{G}}{C_{\Sigma}}V_{G} , \nonumber \\
            \Delta E_{2}^{\pm} =\Delta U^{\pm} \pm
            \frac{eC_1}{C_{\Sigma}}V_D \mp \frac{eC_{G}}{C_{\Sigma}}V_{G},
\label{etr}
\end{eqnarray}
where $C_{1,2,G}$ and $C_{\Sigma}$ are the terminal and total
capacitances respectively, $V_D$ and $V_G$ are drain and gate voltages respectively, and $\Delta U^{\pm}$ is the Coulomb offset
for the addition or removal of an electron. Notice that we now have a
simpler analytical and closed form solution to our set of master
equations, using our trick of summing over excitations. In doing so,
we also brought in simpler circuit parameters such as junction
resistance and capacitance that aid in our analytical understanding
of threshold voltages and current magnitudes.

When considering the ``orthodox'' Coulomb Blockade theory in the
regime of strong contact asymmetry ($R_2 \gg R_1$), it is
illuminating - and not overly limiting - to consider what happens at
very low temperatures. Following the analysis of Hanna and Tinkham
\cite{Han91}, at low temperatures the ensemble distribution of
electrons on the middle electrode can be described by a delta
function $\delta_{n,n_0},$ where $n_0$ is the most probable number
of electrons. The signs of $n_0$ and $Q_0$ in the following
equations will differ from \cite{Han91} because we are using a
negative charge carrier convention. The delta function probability
density reduces Eq.~(\ref{eq:curr_master}) to $I(V_D,V_G) =
e/\hbar[R^{(2)}_{n_{0}\rightarrow{n_0\mp1}}-R^{(2)}_{n_{0}\rightarrow{n_0\pm1}}]$.
For low bias the Coulomb cost of electrons tunneling across the
contacts is high, resulting in a zero-conductance region limited by
the positive and negative threshold voltages $V_{CB}^{(+)}$ and
$V_{CB}^{(-)}$, respectively. Outside of this region the transition
rates simplify:
\begin{eqnarray}
 I(V_D,V_G) = & \displaystyle\frac{1}{R_{2}C_{\Sigma}}[&-(n_{0}e-Q_0)+C_{1}V_{D}-C_{G}V_{G}\nonumber\\
  & & -\frac{e}{2}sgn(V_{D}-V_{CB}^{(-)})]\mbox{ ,}
\label{eq:approxIV}
\end{eqnarray}
where $sgn$ denotes the Heaviside sign function. The linearity of
Eq.~\eqref{eq:approxIV} with drain voltage is only interrupted when
new levels enter into the bias window, causing $n_0$ to change by
$\pm1$, which in turn causes the current to``jump" in value.

\begin{figure}
\centerline{\epsfig{figure=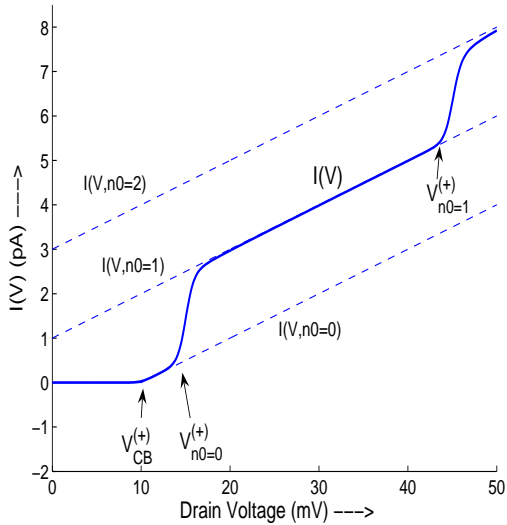,width=3.0in,height=3.0in}}
\caption{(Color Online) Illustration of Orthodox Model:  The
orthodox theory parameters define a set of $I(V,n_0)$ curves,
represented by the dashed lines.  Generally $n_0$ is initially zero.
As a voltage is applied, the current follows the $I(V,n_0=0)$ curve
until $n_0$ changes.  If $|V_{CB}|<|V_{n_0=0}|$, as is the case
here, the current will rise linearly out of the zero-conductance
region; otherwise, there will be a ``jump onset.''  For this
simulation, $R_{1}=10$ M$\Omega$, $R_{2}=6$ G$\Omega$, $C_{1}=8$ aF,
$C_{2}=5.3$ aF, $Q_{0}=0$, and $T=2$ K.} \label{fig4}
\end{figure}

  We thus have an intuitive picture of how I-V ($I-V_D$) curves are constructed
in the orthodox theory.  For given system parameters (R, C, etc.)
and gate voltage, there is a set of $I(V,n_0)$ curves for different
values of $n_0$, as dictated by Eq.~\eqref{eq:approxIV}. The dashed
lines in Fig.~\ref{fig4} correspond to three members of such a set.
In general, both $n_0$ and $I$ are initially zero.  As a drain
voltage is applied, I(V) remains on the $I(V,n_{0}=0)$ curve until
$n_0$ changes, at which point I(V) jumps to the $I(V,n_{0}=\pm1)$
curve. Generalizations of Eqs.~(5,6a) from Hanna and Tinkham
\cite{Han91} allow one to specify the Coulomb Blockade threshold
voltages,
\begin{subequations}
    \begin{align}
       V_{CB}^{(+)} = C_{1}^{-1}(e/2-n_{0}e+Q_0+C_{G}V_{G})\\
       V_{CB}^{(-)} = C_{1}^{-1}(-e/2-n_{0}e+Q_0+C_{G}V_{G})\mbox{,}
    \end{align}
    \label{vcbonsets}
\end{subequations}
and the voltages at which the system transitions from $n_0$ to
$n_{0} \pm 1$ electrons,
\begin{subequations}
    \begin{align}
        V_{n0}^{(+)}=C_{2}^{-1}(e/2+n_{0}e-Q_0-C_{G}V_{G})\\
        V_{n0}^{(-)}=C_{2}^{-1}(-e/2+n_{0}e-Q_0-C_{G}V_{G})\mbox{.}
    \end{align}
    \label{vn0onsets}
\end{subequations}
In Eqs.~(\ref{vcbonsets},\ref{vn0onsets}) the positive superscripts
refer to positive onset voltages, and the negative superscripts are
similarly defined. In Fig.~\ref{fig4}, for example, we can see that
the CB threshold voltage is reached (at $10$ mV) before $n_0$
increases (at $15$ mV), resulting in a linear onset of the current.
If, however, $|V_{n0}^{(\pm)}|$ were smaller than
$|V_{CB}^{(\pm)}|$, then there would be a ``jump'' onset at the
zero-conductance region threshold.

Let us now compare the orthodox and Fock space model approaches to
transport in the CB regime and apply them within the context of
experimental trends.

\section{CB Asymmetries: Gate Dependent Rectification}
One of the simplest consequences of asymmetric contact coupling is
rectification; in other words, a bias direction dependence in the
I-V characteristics. To calibrate with experiments, we not only
concern ourselves with rectification {\it{per-se}}, but also how it
is influenced by a gate. In fact, experiments \cite{jpark,rscott}
showcase gate dependences of the rectification properties that are
arguably more interesting than the rectifications themselves. These
experiments (see for example \cite{jpark,rscott}) show the following
gate-able features: (i) a gate-dependent shift of conductance peak
onsets and (ii) a gate-dependent modulation of the corresponding
conductance peak heights. In addition, there is (iii) a prominent
exchange in conductance peak asymmetry for gate voltage variations
about the charge degeneracy point in the stability diagram
\cite{rscott}. We will argue that much of the relevant physics has
to do with the way the molecule accesses various electronic
excitations under bias, which would require going beyond our
one-orbital model to a multi-orbital system. Charge addition or
removal causes jumps in the I-V, while charge redistribution
(excitation) leads to closely spaced plateaus that merge onto a
linear ramp when summed incoherently. In the rest of the section, we
will explain how each CB model (Fock-space and orthodox)
successfully captures the gate modulation of the asymmetric I-Vs, as
summarized schematically in Figs.~\ref{fig:CB1} and \ref{fig:CB2}.

\subsection{Gate Modulation of current onsets and heights}

\begin{figure}
\centerline{\epsfig{figure=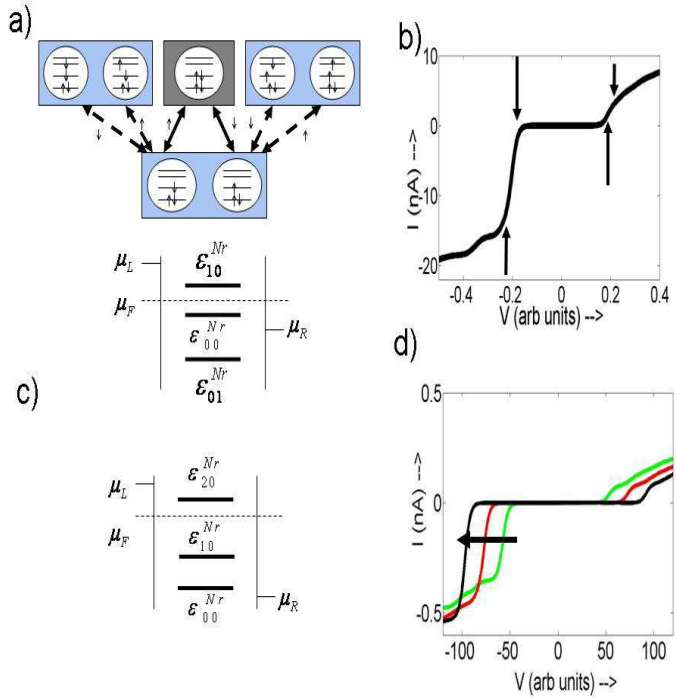,width=3.8in,height=3.8in}}
\caption{(Color Online) Origin of peak asymmetry and variation with gate voltage.
(a) For gate voltages that place the contact Fermi energy $\mu_F$ in
the $N$ electron blockade region, the levels align such that $\mu_F
> \epsilon^{Nr}_{00}$. A state transition diagram (b) shows the
addition and removal of up (down) spins resulting in transitions
$\epsilon^{Nr}_{00}$ (bold double arrow) between ground states of
neutral (light orange) and positively charged species (light blue).
Also shown in state transition diagram are transitions (dashed
double arrow) between various configurations of neutral excited
state (deep orange) and positively charged ground state, labeled
$\epsilon^{Nr}_{10}$. (c) The resulting G-V shows clear asymmetry in
conductance peak height due to there being more ways to add an
electron. Increasing gate bias increases the number of excitations
available, giving a pronounced peak modulation with gate voltage.
The inset shows the corresponding I-V characteristics
\cite{rbhasko,jpark,rscott}} \label{fig:CB1}
\end{figure}

The onset of conduction is determined by the offset between the
equilibrium Fermi energy and the first accessible transition energy,
marked $\epsilon^{Nr}_{00}$ in Fig.~\ref{fig:CB1}
(following the nomenclature in section II A).
This can be varied by varying the gate voltage, thereby accounting
for the variation in conductance gap with gate bias (Fig.~\ref{fig:CB1}c). While the
current step and corresponding conductance peak are generated by this
threshold transition, there follows a quasi-ohmic rise in current
leading to a subsequent constant non-zero conductance in the G-V.
This feature arises from the sequential access of several closely spaced
transport channels under bias, arising
from excitations within the $N$ and $N-1$ electron subspaces \cite{rbhasko}.
While net charge addition and removal come at large Coulomb prices, excitations
involve charge reorganization within the Fock space that cost much smaller
correlation energies.

The presence of multiple orbitals generates several configurations
of excited states, creating more accessible transport channels
within the bias window. For example, in Fig.~\ref{fig:CB1}(a)
conduction occurs simultaneously via the $\epsilon^{Nr}_{10}$ and
$\epsilon^{Nr}_{00}$ removal channels. $\epsilon^{Nr}_{10}$
corresponds to a transition between the first excited state `1' of
the $N$-electron neutral species, and the ground state `0' of the
$N-1$ electron cationic species. We show four possible
configurations corresponding to the transport channel
$\epsilon^{Nr}_{10}$ and the corresponding I-V
(Fig.~\ref{fig:CB1}b). Increasing the gate bias increases both the
threshold for current conduction and the number of such excited
state channels accessed by the contacts, thereby altering the height
of the corresponding conductance peak with gate bias
(Fig.~\ref{fig:CB1}c).

The previous paragraph illustrates the origin of gate-modulated
current as rationalized by the {\it{Fock-space CB model}}. One can
also explain this within the {\it{simpler orthodox model}}, which
ignores the identities of the resolved excitations by incoherently
summing over them. Under the approximations of contact asymmetry and
low temperature, the rate $R^{(j)}_{N\rightarrow{N\pm1}}$ is linear
in the transition energies $\Delta E_{j}^{\pm}$ that increase with
drain voltage. With increasing gate bias one needs a larger
corresponding drain bias to overcome the zero-conductance regime. At
this higher drain voltage the coupling has a greater value and,
consequently, the current magnitude is larger. Physically, the drain
voltage-dependence of the coupling represents a linear approximation
of the excitation spectra.  Even though the orthodox model
indiscriminately sums the excitations within the $N$ and $N-1$
subspaces, the fact that it captures them at all allows it to
qualitatively capture the modulation of current height.

\begin{figure}
\centerline{\epsfig{figure=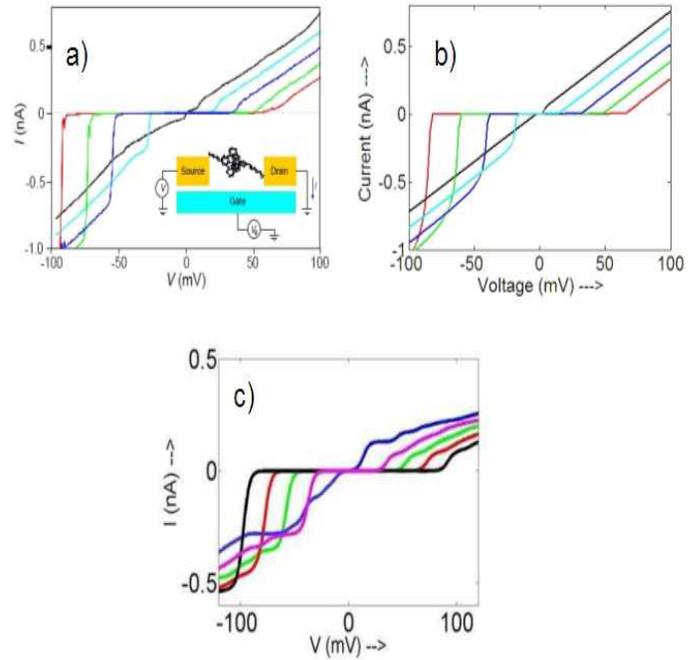,width=3.8in,height=3.8in}}
\caption{(Color Online) Experiment from J. Park \textit{el al.}
\cite{jpark} showing gate-rectification properties in the Coulomb
Blockade regime: a) Experimental traces, b) Orthodox fit with
parameters $C_1 = 0.624$ aF, $C_2 = 0.486$ aF, $C_G = 0.0708$ aF,
$R_1 = 1$ M$\Omega$, $R_2 = 75$ M$\Omega$, $Q_0=-0.05e$ and $T=2.2$
K. c) Fit from Fock-space model.} \label{fig:fig7}
\end{figure}

Figs.~\ref{fig:fig7}(a) and \ref{fig:Zhitenev}(a) show experimental
evidence of gate-modulation of current onsets and heights.  Fig.
\ref{fig:fig7}(a), from J. Park \textit{et al.}, shows a simple
experiment for which the negative bias onset is set by moving from
the $N$ to $N-1$ electron subspace, while the positive bias onset
starts where the $N$ electron excitation spectrum moves into the
bias window. Figs.~\ref{fig:fig7}(b,c) show the abilities of both
the orthodox model and the Fock-space model to capture the
gate-modulated features of the experiment.
Fig.~\ref{fig:Zhitenev}(a), by Zhitenev \textit{et al.}, is a
similar experiment with slightly more complex features. Curves (d-h)
show the same negative bias onset as seen in Fig.~\ref{fig:fig7},
but as the gate voltage is further decreased, the negative onset
changes to a linear onset, representing access to the $N$ electron
excitation spectrum.  The positive bias shows that decreasing gate
voltage brings the $N+1$ level closer to the bias window, and for
curves (a-c) the positive bias type consequently becomes a ``jump''
onset. In spite of these new degrees of freedom that must be
captured, the orthodox theory still models very accurately, as seen
in Fig.~\ref{fig:Zhitenev}(b). It is worth noting that the x-axis in
Fig.~\ref{fig:Zhitenev} is $V_{SD}$, rather than $V_{DS}$, which
must be accounted for when using Eqs.~(\ref{gamma}-\ref{vn0onsets})
of the orthodox theory. A discussion of the extraction of orthodox
parameters from the experimental curves in Figs.~\ref{fig:fig7} and
\ref{fig:Zhitenev} is included in the appendix.

Mathematically, it is straightfoward to understand the dependence of
current height on gate voltage within the orthodox theory.  Using
the experiment of J. Park. \textit{et al.} (Fig.~\ref{fig:fig7}a) as
an example, we see that there is a jump onset for negative bias
voltages. Therefore, $\left|V_{n0}^{(-)}\right| <
\left|V_{CB}^{(-)}\right|$. At $V_{n0}^{(-)}$, the I-V transitions
to the $I(V,n0=-1)$ curve, so to find the current height at the
onset voltage we can use Eq.~\eqref{eq:approxIV} for $n_0=-1$.
Inserting $V_{n0}^{(-)}$ for $V_D$, one finds:
\begin{equation}
I(V_{n0}^{-})= -\frac{C_{1}+C_{2}}{R_{2}C_{2}C_{\Sigma}} \left(Q_0
+e/2+C_{G}V_{G} \right)
\end{equation}
Clearly the magnitude of the current at the onset voltage increases
with gate voltage (Fig.~\ref{fig:fig7}b), matching the experimental
result seen in Fig.~\ref{fig:fig7}(a).

\begin{figure}
\begin{center}
    \mbox{
      \subfigure{{\epsfig{figure=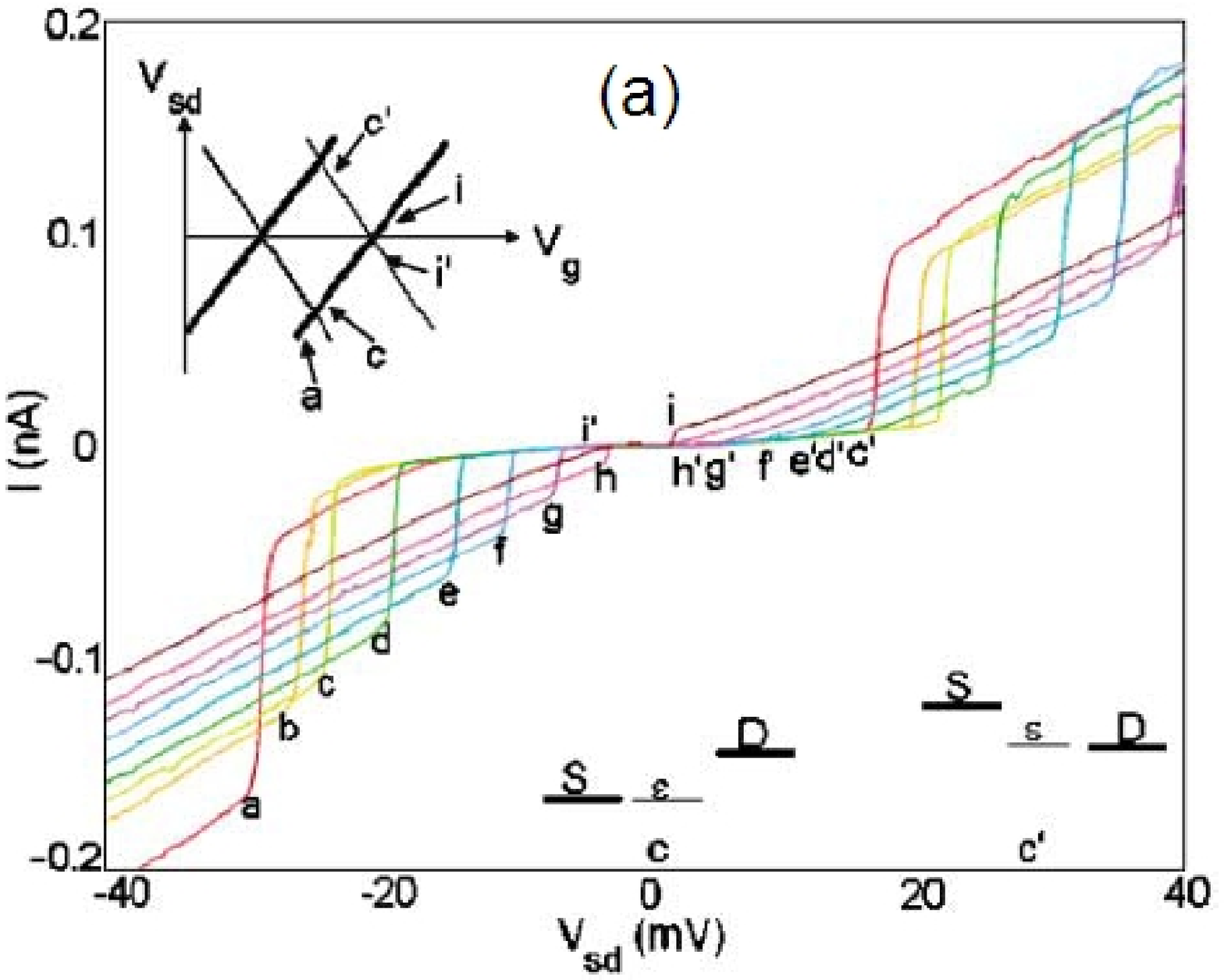,width=1.6in,height=1.5in}}} \quad
      \subfigure{{\epsfig{figure=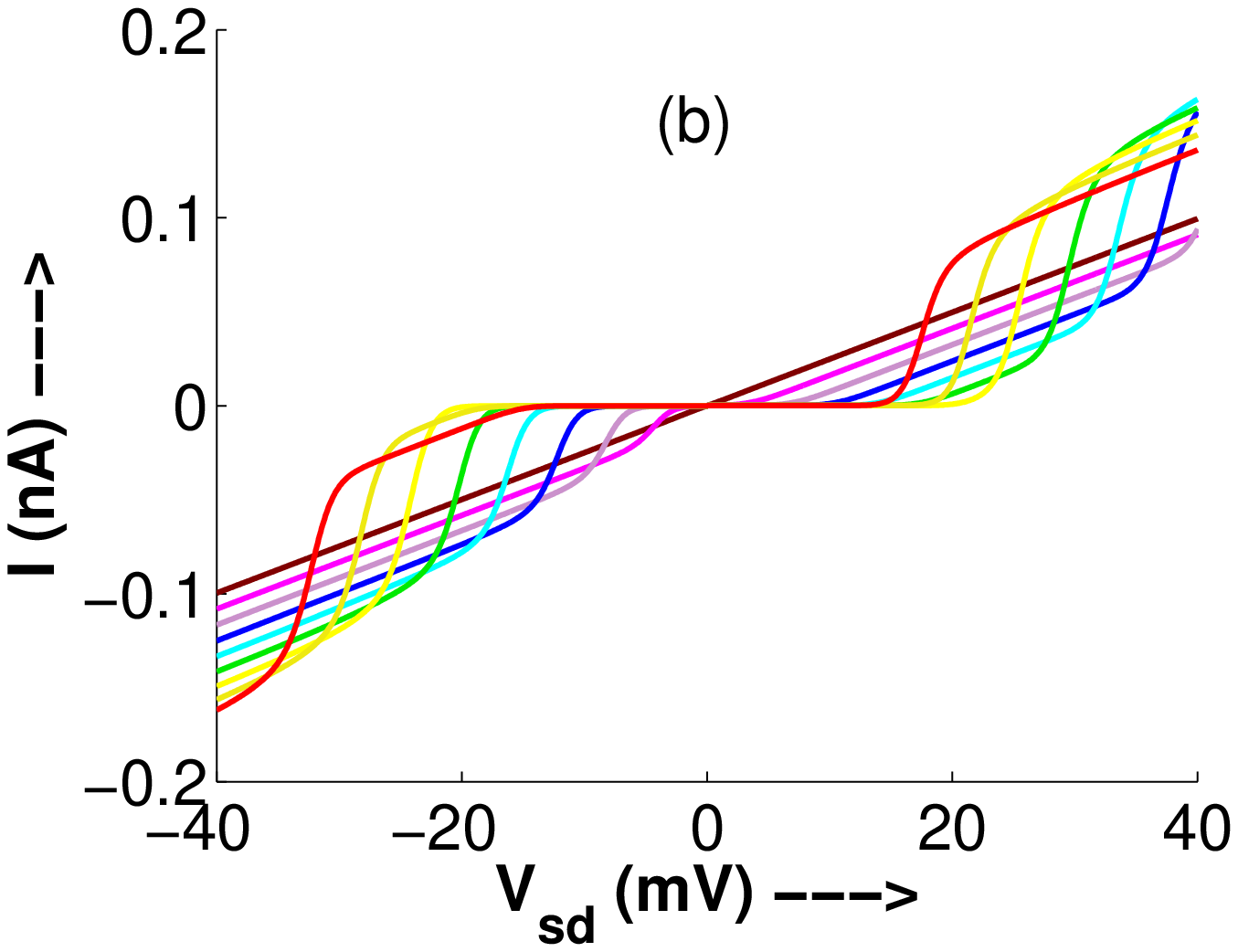,width=1.6in,height=1.5in}}} \quad
      }
\caption{(Color Online) Asymmetric CB results showing (a) experiment
\cite{zhitenev} and (b) orthodox theory with parameters $C_1=3.70$
aF, $C_2 = 3.24$ aF, $C_G=0.061$ aF, $R_1=2$ M$\Omega$, $R_2 = 210$
M$\Omega$, $Q_0=0.175e$, and $T=4.2$ K.} \label{fig:Zhitenev}
\end{center}
\end{figure}

The accuracy of the orthodox simulation would imply at the very
least that individual excitations do not play an important role in
the transport characteristics seen.  Because the experimental I-V
has a strongly linear dependence on drain voltage seen in
Eq.~(\ref{eq:approxIV}), it seems that the experiments may have
measured transport through a metallic particle, which has a
relatively featureless density of states, as is assumed in the
orthodox model.

\subsection{Peak Exchange}

Experiments exhibit a characteristic flipping of conductance peak
asymmetry around the charge degeneracy point A in the stability
diagram (Figs.~\ref{fig:CB2}a,b). Figs.~\ref{fig:CB2}(e,f) show
typical calculated G-Vs in this regime, featuring conductance peak
asymmetries with respect to voltage bias, arising due to asymmetric
contact couplings ($\gamma_L \gg \gamma_R$). Within the Fock-space
model, this can be explained by enumerating the channels for adding
and removing electrons under bias (Figs.~\ref{fig:CB2}b,d), with the
weaker right contact once again setting the rate limiting step. The
dominant transport channel $\epsilon^{Nr}_{00}$ corresponds to
electronic transitions between the neutral and cationic ground
states \cite{rbhasko}, states which SCF theories do take into
account. In the CB limit, however, there are additional electronic
excitations that are accessible with very little Coulomb cost. These
states are responsible for the peak asymmetry exchange observed in
these experiments, as we will now explain.

The origin of this asymmetry can be understood with a simple model
system: in our case, a quantum dot with 8 spin-degenerate levels and
$N=4$ electrons in its ground state. When the Fermi energy lies to
the immediate right of the charge degeneracy point as shown in
Fig.~\ref{fig:CB2}(a), only transitions between the $N$ and $N-1$
electron states (4 and 3) are allowed, with the weaker right contact
setting the rate-limiting step. For positive bias on the right
contact an electron can be removed from the 4-electron to the
3-electron ground state in two ways (Fig.~\ref{fig:CB2}b)). For
negative bias, however, the electron removed by the left contact can
be replenished by the right contact back into the 4-electron ground
state, but also into one of many possible excited states
$\epsilon^{4r}_{i0}=E^{4}_{i}- E^{3}_{0}$, (i $>$ 0). Since there
are more ways to bring the electron back (6 shown here), the
conductance is larger for negative bias (Fig.~\ref{fig:CB2}e). The
situation changes dramatically for a different position of the Fermi
energy (Fig.~\ref{fig:CB2}c) in the stability diagram lying to the
left of the charge degeneracy point A with three electrons at
equilibrium. For negative bias the right contact adds an electron
from the 3 to the 4-electron ground state, while for negative bias
it returns it to the $j^{th}$ 3-electron excited state through
transitions $\epsilon^{4r}_{0j}=E^{4}_{0}-E^{3}_{j}$. There now are
more ways to remove than add charge (Fig.~\ref{fig:CB2}d), so the
asymmetry flips (Fig.~\ref{fig:CB2}f).

\begin{figure}
\centerline{\epsfig{figure=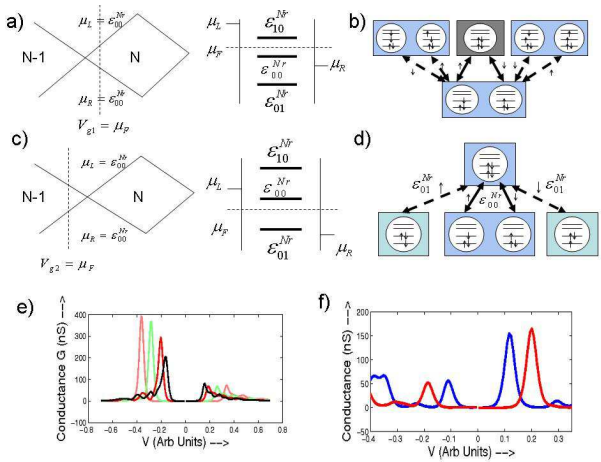,width=3.2in}}
\caption{(Color Online) Origin of exchange in asymmetry of conductance peaks: (a),
(b) Schematic of vicinity of charge degeneracy point A in Coulomb
diamond. Also shown are the corresponding energy diagrams at
threshold. Notice a different set of threshold transport channels
between (a) and (b). (c) and (d) State transition diagrams illustrating
the different excitation spectra accessed on either side of the
charge degeneracy point A. (e) and (f) G-V plots for scenarios (a) and
(b) Notice the clear peak-exchange as a result of accessing a
different excitation spectra in either case.} \label{fig:CB2}
\end{figure}

Analogous to the {\it{Fock-space model}}, the {\it{orthodox model}}
also captures gate-dependent peak exchange, in spite of its
approximate treatment of excitations.  The origin of the asymmetry
is once again transitions between the $N$ to $N-1$ electron regimes.
In Fig.~\ref{fig4}, we can see that such a change results in a jump
onset that has a much higher conductance value than a linear onset.
Moving from the $I(V,n0)$ curve to the $I(V,n0-1)$ curve, therefore,
essentially captures the excitations of the $N-1$ electron spectrum
that are pivotal to the argument in the preceding paragraph.  The
fact that the conductance peak switches across zero bias only means
that the zero-bias state of the system changes from $N$ to $N-1$
electrons, which the orthodox method clearly captures.

Peak exchange has been reported experimentally \cite{rscott}, as
seen in Fig.~\ref{fig:fig6}(a).  Fig.~\ref{fig:fig6}(b) shows an
orthodox simulation of the experiment.  One can see close
qualitative as well as quantitative agreement between the
experimental data and the theoretical simulation.  The conductance
peaks have similar magnitudes, and the exchange of the peak
asymmetry occurs at $-3.75$ $V$ in both graphs. The evident validity
of the orthodox theory in this case demonstrates its ability to
capture excitation features, as long as the features can be linearly
approximated.

\begin{figure}
\centerline{\epsfig{figure=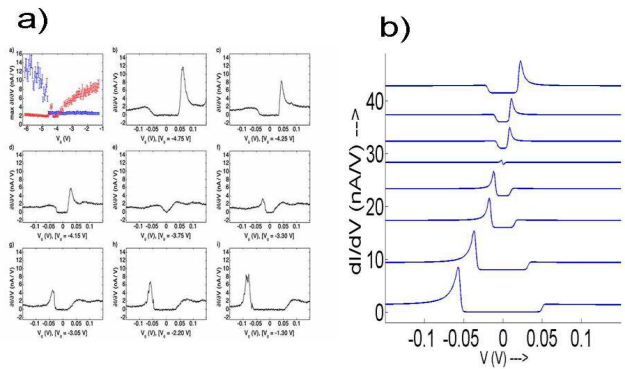,width=3.5in,height=2.0in}}
\caption{(Color Online) (a) Experimental trace demonstrating peak
asymmetry exchange \cite{rscott}. Compare with Fock-space model
results in Fig.~\ref{fig:CB2}(e,f). (b) Orthodox simulation
reproducing peak exchange observed in the experiments. The
parameters are: $R_{1}=35$ M$\Omega$, $R_{2}=350$ M$\Omega$,
$C_{1}=0.673$ aF, $C_{2}=0.612$ aF, $C_{G}=0.0135$ aF,
$Q_{0}=-0.18e$, and $T=4.2$ K.} \label{fig:fig6}
\end{figure}

\section{Limitations of the Orthodox Model} In the previous sections,
we saw that the orthodox model has been fairly successful in
reproducing the three key trends in asymmetric CB transport; namely,
(i) rectification, (ii) gate-modulation of rectification and (iii)
exchange of rectification. The agreement between experiment and
theory are quantitatively quite close, indicating that the
Fock-space model's handling of well-resolved excitations is perhaps
an overkill, especially considering its substantially greater
computational complexity. It is tempting to conclude that molecules
with redox-active centers only exhibit incoherent sums of
excitations that do not manifest well-resolved features. In this
section, we point out examples where {\it{the discrete excitation
spectrum can indeed play a noticeable role in molecular transport
experiments, making an orthodox theoretical treatment quite
inadequate}}.

Molecules are unique in that they can potentially exhibit both
charge and size quantization \cite{rbhaskob}. Charge quantization is
enforced by $U
> \Gamma$, which amounts to a large contact resistance compared to
the resistance quantum (the correspondence arises by relating the broadening
$\Gamma$ to an RC time-constant through the uncertainty principle,
using the capacitance $C$ to set the single-electron charging energy $U$). At the same time, the
small sizes of molecules make their spectra discrete. While transport
examples showcasing discrete molecular
signatures are relatively rare, they are pretty commonplace in the
literature of inorganic quantum dots or `artificial molecules' \cite{rbanin}.
Part of the reason is the lower broadening in these dots, both from
the contact as well as incoherent scattering (which is larger in
organic molecules owing to their conformational flexibility).
Recently, a novel negative differential resistance (NDR) has been
reported in inorganic double quantum dots \cite{tar}, that has been
explained via the formation of an excited triplet state
\cite{tar,rbhasko2}. Other quantum dot experiments routinely show
Coulomb diamonds with very well resolvable excitation lines
\cite{sampaz} such as due to cotunneling \cite{cotunel} and Kondo correlations \cite{kondo}.

Well resolved excitation lines in the conductance
spectra manifest as varying plateaus in the I-V characteristics.
Consider the low bias ($V_D <500$ mV) I-V characteristics of the
plot from J-O. Lee  {\it{et al.}} \cite{rdekker}, seen in
Fig.~\ref{fig:fig8}. A striking feature is the existence of a plateau
at onset followed by several other plateaus, sometimes
even merging into a quasi ohmic rise as a result of several
unresolvable plateaus. This feature, observed in multiple experiments from
other experimental groups, can be easily explained within the
Fock-space model by keeping track of individual molecular excitations.
For example, it is well known that the gap between ground and first
excited states, involving charge addition or removal, is greater than
the gap between subsequent excitations involving charge reorganization.
In such a case it is clearly seen that a brief plateau occurs at
threshold that persists until the first excitation is been accessed,
as discussed in detail in \cite{rbhasko}.

This leads to an important
point. Normally small plateaus (few tens of mV) are naturally associated
with vibronic modes whose energy scales lie in that range. Coulomb
interactions are usually assumed to be larger in energy, because the
energy to add or remove an electron is much larger. However, the correlation
energy to reorganize charges could actually be
very small, are therefore quite capable of explaining
a range of plateau widths seen experimentally, as our calculations show.

The orthodox theory, on the other hand, cannot even qualitatively
match the experimental data in Fig.~\ref{fig:fig8}, owing to its
{\it{inability to incorporate size quantization effects and the associated
discrete spectra}}.  From Eq.~\ref{eq:approxIV}, it is clear
that outside of the zero-conductance region and excluding jumps due
to changes in $n_0$, conductance values in orthodox theory must
{{remain constant}}, with a value of $R_{1}/R_{2}C_{\Sigma}$. {\it{The
orthodox theory can capture a plateau, and it can also capture a
linear rise, but it does not seem to capture both in the same I-V curve.}}
Fig.~\ref{fig:fig8}(c) shows the best attempt at modeling the experimental
data in Fig.~\ref{fig:fig8}(a) within orthodox theory; {{the plateau
followed by a linear rise seems hard to duplicate}}.

\begin{figure}
\centerline{\epsfig{figure=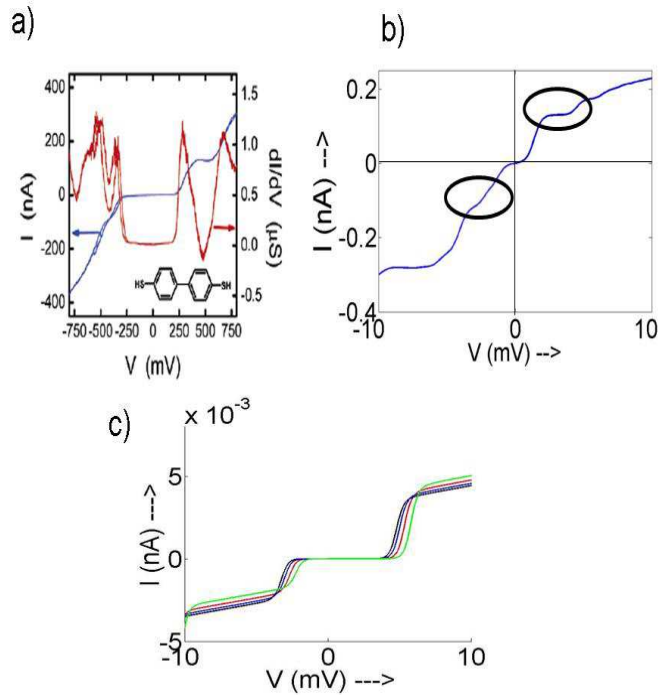,width=3.8in,height=3.8in}}
\caption{(Color Online) Limitation of Orthodox Model: a) Experimental trace \cite{rdekker} showing fine structure. b) 
The fine structure in I-V's, result from keeping track of excitations explicitly within
the Fock-space CB model. c) An orthodox calculation merely maintains the same slope between charge
addition jumps, thus may not reproduce any fine structure.}
\label{fig:fig8}
\end{figure}

A second limitation of the orthodox model comes in its
{\it{treatment of gate voltage}}. Even though it effectively modeled
the data in Figs.~\ref{fig:fig7}, \ref{fig:Zhitenev} and
\ref{fig:fig6}, there again exist experimental features that the
orthodox theory cannot even qualitatively model.  Looking at Eqs.
\eqref{vcbonsets} and \eqref{vn0onsets} one can see that a change in
gate voltage causes a {\it{translation}} in the Coulomb Blockade
threshold voltages; a similar effect, albeit in the opposite
direction, is seen for the voltage limits at which $n_0$ changes.
Fig.~\ref{fig:fig5} shows how an orthodox I-V curve changes with
gate voltage for the four possible onset combinations - which come
from having linear or jump onsets at positive and negative bias. The
scaling of the conductance gap in Figs.~\ref{fig:fig5}(a,b)
successfully explained the experiments in Figs.~\ref{fig:fig7} and
\ref{fig:fig6}. However, for experiments with symmetric onsets
(Figs.~\ref{fig:fig5}c,d), one can see that the orthodox theory
predicts an overall translation in the I-V curve with gate voltage.
One would think that changing the gate voltage would only shift the
conducting level closer to or further from the bias window, thereby
narrowing or widening the I-V curve, respectively. Indeed,
experimental evidence demonstrates such narrowing \cite{rralph2},
and the Fock-space model captures that quite easily. The orthodox
theory was unable to match this experimental trend even
qualitatively.

\begin{figure}
\centerline{\epsfig{figure=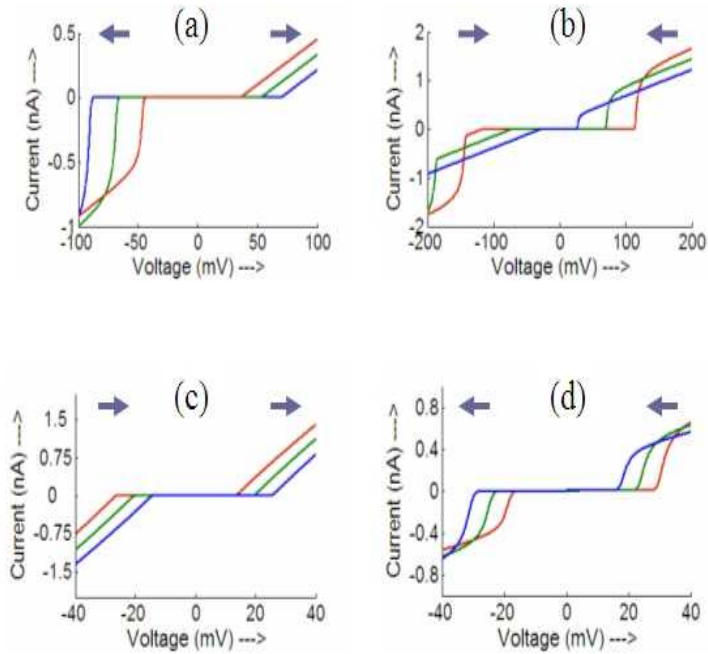,width=3.8in,height=3.8in}}
\caption{(Color Online) Gate-voltage dependence of I-V curves in the orthodox
theory.  Blue arrows indicate the direction of movement of the onset voltages
with increasing gate voltage.  The four plots correspond to the four
possible onset combinations: (a) a jump onset at negative bias and a
linear onset at positive bias, for which the I-V conductance
gap {\it{widens}} with increasing gate voltage; (b) a linear onset then
a jump onset, with the gap {\it{narrowing}} around zero-bias; (c)
two linear onsets, which corresponds to a {\it{translation}} of the I-V
curve, and (d) two jump onsets, which correspond to a {\it{translation}}
in the opposite direction of (c).} \label{fig:fig5}
\end{figure}

\section{Discussions}
It may seem that the proper treatment of well-resolved excitations
is rather academic with respect to molecular experiments. However,
we believe that there can be more important experimental features
that need a proper quantitative theory as transport spectroscopy of
molecules under becomes feasible \cite{ucla}. In fact, a possible
explanation for molecular NDRs \cite{tour,kiehl} could necessitate
keeping track of excitations in a donor-acceptor molecular system
\cite{rbhasko2}; in particular, the notably different lifetimes of
ground and excited states, analogous to issues relevant to the
double quantum dot literature. Small molecules could function as
tunable quantum dots with high single-electron charging energies.
Molecular quantum dots coupled to transistor channels can be
important for the detection, characterization and manipulation of
individual spin qubits \cite{ucla}, the transistor conductance
providing a way to achieve electronic read-out \cite{rabi}. The
large charging energies could allow redox-active molecules to
operate as storage centers for memory \cite{misra}. Finally,
rectification is important to avoid parasitic pathways in cross-bar
logic based architectures \cite{stan}.

The accurate treatment of well-resolved excitations is crucial to
molecular operation in this scattering regime. But the price paid
for this accuracy is the loss of simplicity associated with orthodox
theory. Instead, we will need a major improvement in computational
algorithms to handle the exponential scaling of the many-body Fock
space, such as partial configuration interaction (CI) with frontier
orbitals coupled to the standard master equations. Major challenges
involve the proper inclusion of the contact boundary conditions, and
doing justice to broadening \cite{rWacker}. Also, one could
incoherently sum selected excitations and resolve the relevant ones,
if we have some prior idea which ones could be important. {\it{Spin
unrestricted approaches may capture some of the salient effects,
although they are unlikely to capture the current heights (which
depend on difficult parameter-independent combinatorial arguments),
and even more significantly, the number of current plateaus (since
an energy-independent one-electron potential does not generate
enough poles).}} Needless to say, there is enormous room for
theoretical activity and device potential in this domain.

It is worth emphasizing that the wide success and popularity of the
SCF-NEGF approach often leads researchers to automatically assume
that the method will work in the CB regime \cite{revers,rpal}. The
NEGF method is indeed prolific in its flexibility to incorporate
contact microstructure, molecular chemistry, broadening,
intereference and electrostatic details. However, the method, as
widely implemented, is fundamentally limited by the need to work
with a one-electron potential, at best incorporating many-body
effects as corrections to this potential. While this may be
sufficient to capture equilibrium properties like total energy,
transport measurements can potentially probe the rich excitation
spectra that are hard to incorporate readily into the average
electronic potential, even parametrically. For a typical
self-energy, there simply are not enough poles in the one-electron
Green's function to do justice to the many more many-electron
transitions characterizing transport, not to mention the
complications of the full counting statistics in Fock space under
non-equilibrium that show up as effective
{\it{interaction-independent}} degeneracies of the open-shell
levels.

In this paper, we have discussed contact asymmetry as a paradigm to
delve into various experimental features, outlining the qualitative
difference and cross-over between the weakly correlated SCF and the
strongly correlated CB regimes. An understanding of the asymmetry
was provided using a basic model, and then extended to conjugated
molecular systems. Novel experimental-trends were identified in the
CB-regime of transport. Two different approaches, the Fock-space CB
and the orthodox CB, were discussed in detail. While a simpler
orthodox approach captures the salient asymmetric effects, the
Fock-space approach may be essential in examples where the interplay
between specific excitations governs the principle operating physics
and possibly also interesting device applications envisaged.

\section{Acknowledgments}
We would like to thank S. Datta, L. Harriott, G. Scott and H-W.
Jiang for useful discussions. This project was supported by the
DARPA-AFOSR grant.

\section{Appendix I: Extracting parameters for orthodox model}
The low temperature, strong contact asymmetry approximation to the
orthodox theory, as introduced in section III, is most useful
because it simplifies the process of simulating experimental data.
Whereas the general orthodox formalism requires best-fit
computational techniques, Eqs.~(\ref{eq:approxIV}-12) allow for a
straightforward analytic calculation of the system circuit
parameters.  Consider Fig.~(\ref{fig:fig7}) as an example of data to
be simulated with the orthodox theory.  To start, consider the $V_G
= -0.85 V$ I-V curve (light blue).  We have a linear onset at
approximately $19$ $mV$, and a ``jump'' onset at, say, $-24$ $mV$.
Plugging these values into Eqs. (11a) and (12b), respectively, and
solving finds:
\begin{eqnarray}
C_{1}(19mV)=C_{2}(24mV) \mbox{, and} \\
C_1=\frac{e+2Q_{0}+2C_{G}V_{G}}{2(19mV)} \mbox{.}
\end{eqnarray}
Consideration of a second I-V curve in the data gives us further
information; here we will use the $V_G=-0.55 V$ (green) curve, which
has a positive onset voltage of approximately $53$ mV. After
plugging the correct onset and gate voltage values into Eq. (11a)
for both I-V curves considered, one can solve for a new, independent
equation:
\begin{equation}
C_{1}=\frac{\Delta V_{G}}{(53-19)mV}C_{G}=8.82C_{G}
\end{equation}
We now have three equations and four unknowns ($C_1$, $C_2$, $C_G$,
and $Q_0$).  There are now two approaches one can take.  The first,
which was applied in the simulation shown in Fig. 4(b), is to simply
take a small, reasonable value of $Q_0$, such as $-0.05e$, and solve
the remaining equations for the other variables.  However, this will
not always work, for a reason that is somewhat subtle.  It is
important to note in Fig. 4(a) that not only does the $V_G=-0.85 V$
I-V curve have a linear onset at $19$ $mV$, but it also does not
have a jump onset at positive bias until at least $100$ $mV$.
Setting Eq. (12a) greater than $100$ $mV$ provides a fourth
restriction.  Indeed, for the experiment in 4(a) this is trivially
satisfied; Fig.~\ref{fig:Zhitenev}, on the other hand, is an example
of data for which such a fourth equation is necessary for a
successful simulation.

A final parameter one can obtain is $R_2$.  From
Eq.~\eqref{eq:curr_master} we know that the conductances of the
linear portions of orthodox curves have the value
$C_{1}/R_{2}C_{\Sigma}$.  Having already calculated all of the
capacitance values it is thus straightforward to find the slope of
the experimental I-V curve and find $R_2$.  $R_1$ cannot be found so
easily, but its relative unimportance in the regime of contact
asymmetry means that a simple estimate of $R_{1} \sim R_{2}/100$
will usually suffice.

\end{document}